\documentclass[apsprl,twocolumn,superscriptaddress,longbibliography]{revtex4-1}

\usepackage{graphicx}
\usepackage{amsmath}
\usepackage{ulem}
\usepackage[colorlinks=true,citecolor=blue,linkcolor=magenta]{hyperref}
\usepackage[usenames]{color}
\usepackage{braket}
\usepackage{amssymb}
\usepackage{commath}
\usepackage{csquotes}
\usepackage{dcolumn}
\usepackage{bm}
\usepackage{epstopdf}
\usepackage{lipsum}
\usepackage{subfigure}
\usepackage{siunitx}

\begin{document}

\title{Non-Equilibrium Mass Transport in the 1D Fermi-Hubbard Model}

\author{S.~Scherg}
\affiliation{Fakult\"at f\"ur Physik, Ludwig-Maximilians-Universit\"at M\"unchen, Munich, Germany}
\affiliation{Max-Planck-Institut f\"ur Quantenoptik, 85748 Garching, Germany}

\author{T.~Kohlert}
\affiliation{Fakult\"at f\"ur Physik, Ludwig-Maximilians-Universit\"at M\"unchen, Munich, Germany}
\affiliation{Max-Planck-Institut f\"ur Quantenoptik, 85748 Garching, Germany}

\author{J.~Herbrych}
\affiliation{Department of Physics and Astronomy, University of Tennessee, Knoxville, Tennessee 37996, USA}
\affiliation{Materials Science and Technology Division Oak Ridge National Laboratory, Oak Ridge, Tennessee 37831, USA}

\author{J.~Stolpp} 
\affiliation{Fakult\"at f\"ur Physik, Ludwig-Maximilians-Universit\"at M\"unchen, Munich, Germany}
\affiliation{Institute for Theoretical Physics, Georg-August-Universit\"at G\"ottingen, 37077 G\"ottingen, Germany}
\affiliation{Arnold Sommerfeld Center for Theoretical Physics,  Ludwig-Maximilians-Universit\"at M\"unchen, 80333 Munich, Germany}

\author{P.~Bordia}
\affiliation{Fakult\"at f\"ur Physik, Ludwig-Maximilians-Universit\"at M\"unchen, Munich, Germany}
\affiliation{Max-Planck-Institut f\"ur Quantenoptik, 85748 Garching, Germany}

\author{U.~Schneider}
\affiliation{Fakult\"at f\"ur Physik, Ludwig-Maximilians-Universit\"at M\"unchen, Munich, Germany}
\affiliation{Max-Planck-Institut f\"ur Quantenoptik, 85748 Garching, Germany}
\affiliation{Cavendish Laboratory, University of Cambridge, Cambridge CB3 0HE, UK}

\author{F.~Heidrich-Meisner}
\affiliation{Institute for Theoretical Physics, Georg-August-Universit\"at G\"ottingen, 37077 G\"ottingen, Germany}

\author{I.~Bloch}
\affiliation{Fakult\"at f\"ur Physik, Ludwig-Maximilians-Universit\"at M\"unchen, Munich, Germany}
\affiliation{Max-Planck-Institut f\"ur Quantenoptik, 85748 Garching, Germany}

\author{M.~Aidelsburger}
\affiliation{Fakult\"at f\"ur Physik, Ludwig-Maximilians-Universit\"at M\"unchen, Munich, Germany}
\affiliation{Max-Planck-Institut f\"ur Quantenoptik, 85748 Garching, Germany}

\begin{abstract}
We experimentally and numerically investigate the sudden expansion of fermions in a homogeneous one-dimensional optical lattice. For initial states with an appreciable amount of doublons, we observe a dynamical phase separation between rapidly expanding singlons and slow doublons remaining in the trap center, realizing the key aspect of fermionic quantum distillation in the strongly-interacting limit. For initial states without doublons, we find a reduced interaction dependence of the asymptotic expansion speed compared to bosons, which is explained by the interaction energy produced in the quench. 
\end{abstract}

\maketitle

Many-body physics in one dimension (1D) differs in numerous essential aspects from its higher-dimensional counterparts. Several familiar concepts, such as Fermi-liquid theory~\cite{Baym1991,Giamarchi2004}, are not applicable in 1D. Moreover, many 1D models are integrable, meaning that there exist exact solutions. Examples include the Lieb-Liniger model~\cite{Cazalilla2011}, the Heisenberg chain~\cite{Kluemper} or the 1D Fermi-Hubbard model (FHM)~\cite{Essler-book}. These models exhibit extensive sets of conserved quantities that prevent thermalization~\cite{Rigol2007,Vidmar2016,Kinoshita2006,Hofferberth2007,Gring2012,Langen2015} and can, in lattice systems, lead to anomalous transport properties~\cite{Zotos1997,Heidrich-Meisner2007,Vasseur2016,Karrasch2017}. Cold-atom experiments offer the possibility to study transport properties of strongly-correlated quantum gases in a clean environment. Their excellent controllability enabled far-from-equilibrium experiments~\cite{Ott2004,Strohmaier2007,Schneider2012,Ronzheimer2013,Xia2015} as well as close-to-equilibrium measurements in the linear-response regime~\cite{Xu2016,Anderson2017,Brown2018,Nichols2018} both in extended lattices and mesoscopic systems~\cite{Brantut2012,Valtolina2015,Lebrat2017}. 

\begin{figure}[t!]
        \includegraphics{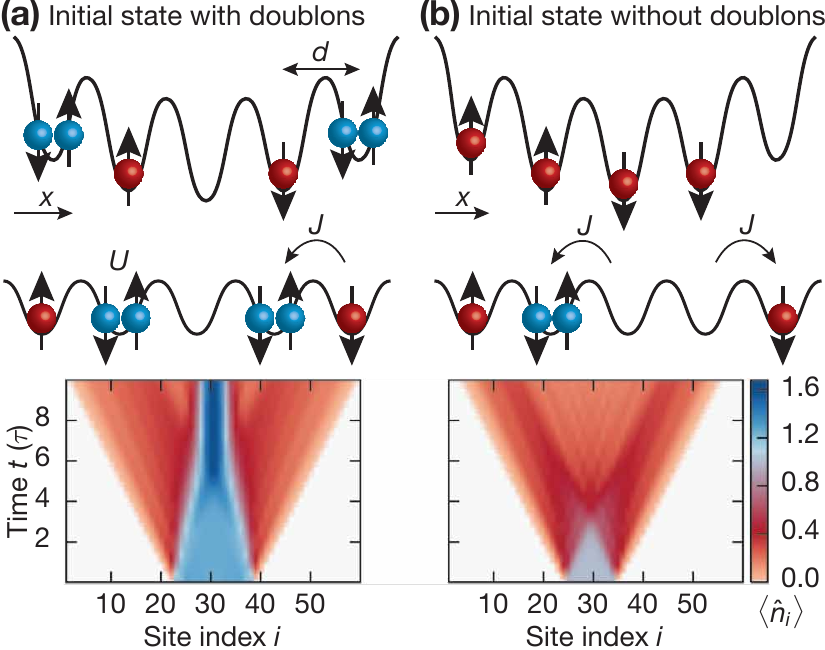}
        \caption{\textbf{Schematics of the expansion experiment.} Top: Initial state of the harmonically trapped two-component Fermi gas with (a) singlons (red) and doublons (blue) and (b) only singlons in an optical lattice. After quenching to lower lattice depths and removing the harmonic trap, fermions expand in a homogeneous 1D lattice; $J = h \cdot 0.54(3)\,$kHz denotes the tunnel coupling, $U$ refers to the on-site interaction strength and $d$ is the lattice constant. The expansion dynamics is dominated by first-order processes: (a) the resonant exchange of singlon and doublon positions leads to quantum distillation, (b) the dynamical formation of doublons results in reduced asymptotic expansion velocities. Bottom: time-dependent density-matrix renormalization group (tDMRG) simulations of the atomic density $\langle \hat n_i \rangle$ for $U=20J$ as a function of time $t$ in units of the tunneling time $\tau=\hbar/J$.}
        \label{fig:schematics}
\end{figure}

Here, we investigate mass transport in the 1D FHM in far-from-equilibrium expansion experiments~\cite{Schneider2012,Ronzheimer2013,Xia2015}, where an initially trapped gas is suddenly released into a homogeneous potential landscape as illustrated in Fig.~\ref{fig:schematics}. There are two distinct regimes of interest in sudden-expansion studies: the asymptotic one, where the expanding gas has become dilute and effectively non-interacting~\cite{Rigol2005,Minguzzi2005,delCampo2006,Jukic2009,Langer2012,Vidmar2013,Campbell2015,Mei2016,Schluenzen2016} and the transient regime, where the dynamical quasi-condensation of hardcore bosons~\cite{Rigol2004,Hen2010,Jreissaty2011,Vidmar2015,Vidmar2017} and quantum distillation~\cite{Heidrich-Meisner2009,Muth2012,Herbrych2017,Xia2015} have been found.

\begin{figure*}[t!]
        \includegraphics{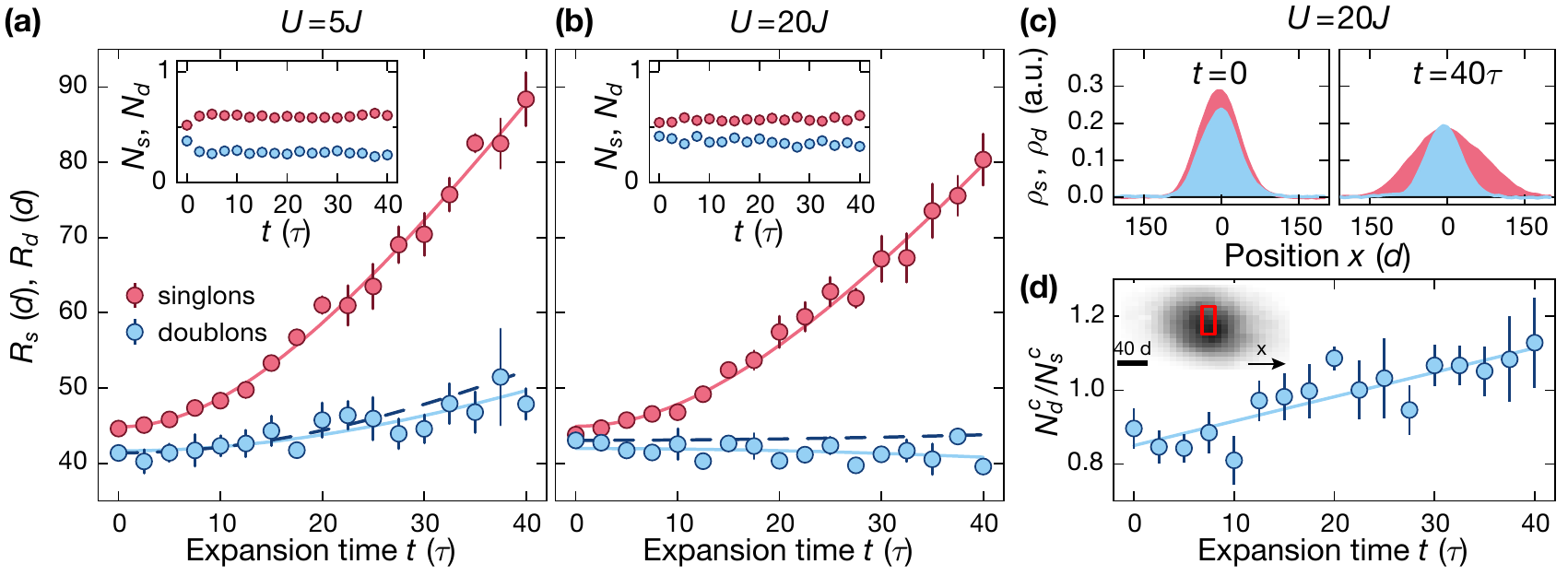}
        \caption{\textbf{Dynamical phase separation of singlons and doublons.} Half-width-at-half-maximum (HWHM) size $R_{s,d}$ of singlon (s, red) and doublon (d, blue) clouds as a function of time for (a) $U=5J$ and (b) $U=20J$.
The dashed lines illustrate the hypothetical expansion of a non-interacting doublon cloud with effective tunneling $J_{\rm eff}$~\cite{suppmat}. Insets: Number of atoms on singly- and doubly-occupied sites, $N_s$ and $N_d$, as a function of time.
(c) Experimental snapshots of the integrated line densities for singlon and doublon clouds, $\rho_s(x)$ and $\rho_d(x)$, at $t=0$ (left) and $t=40\tau$ (right). (d) Ratio $N^c_{d}/N^c_{s}$ of atom numbers on doubly- and singly-occupied sites in the central region of the cloud (red rectangle in the inset) as a function of time for $U=20J$. Every data point is averaged over four measurements and error bars represent the standard-error-of-the-mean. Solid lines are guides to the eye.}
        \label{fig:phase_separation}
\end{figure*}

Quantum distillation occurs for large interactions. It relies on the dynamical demixing of fast singlons (one atom per site) and slow doublons (two atoms per site) during the expansion: while isolated doublons only move with a small effective second-order tunneling matrix element $J_{\rm eff}=2J^2/U \ll J$ for $U\gg J$~\cite{Winkler2006,Rausch2017}, neighboring singlons and doublons can exchange their positions via fast, resonant first-order tunneling processes. Thus, after opening the trap, singlons escape from  regions of the cloud initially occupied by singlons and doublons, leading to a spatial separation of the two components. Without an increase of the doublon density in the central region, this regime is termed weak quantum distillation~\cite{Herbrych2017}. Ideal initial-state conditions can lead to a strong version of quantum distillation, where the spatial separation of singlons and doublons yields a contraction of the doublon cloud radius [Fig.~\ref{fig:schematics}(a)]. In the extreme limit this distillation could be used to purify a finite-temperature band insulator~\cite{Heidrich-Meisner2009}, thereby dynamically generating low-entropy regions. This would represent a major advancement in the ongoing quest of realizing fermionic many-body physics at the lowest entropy scales~\cite{Bernier2009,Ho2009,Mazurenko2017,Chiu2017,McKay2011}. So far, experimental evidence for weak quantum distillation has only been found for bosons~\cite{Xia2015} at intermediate interaction strengths, where doublons can decay into singlons on time scales relevant to quantum distillation. 

In this work, we investigate the non-equilibrium mass transport in the 1D FHM starting from initial product states in deep optical lattices~\cite{Schneider2012,Ronzheimer2013}, while close-to-thermal initial states were used in Ref.~\cite{Xia2015}. The expansion dynamics is initiated by two simultaneous quenches: a sudden increase of the tunnel coupling, resulting in a quench from almost infinite to finite $U/J$, and a sudden removal of the harmonic trap (Fig.~\ref{fig:schematics}). We prepare initial product states with or without doublons (Fig.~\ref{fig:schematics}) and quantitatively investigate the time evolution of singlon and doublon densities individually. For initial states with doublons, we find a distinct dynamical phase separation between singlons and doublons, which is the fundamental mechanism of fermionic quantum distillation [Fig.~\ref{fig:schematics}(a)]. For initial states without doublons [Fig.~\ref{fig:schematics}(b)], we study interaction effects in the asymptotic expansion velocities~\cite{Mei2016}. We observe that the cloud expands rapidly at all interaction strengths with slightly smaller velocities at intermediate values, in agreement with our numerical simulations. This can be interpreted in terms of the interaction energy produced in the quench of $U/J$, which leads to the dynamical formation of doublons~\cite{Ronzheimer2013,Vidmar2013,Heidrich-Meisner2008,Vidmar2013}.

{\it Experiment.} We prepare a degenerate Fermi gas of $30(1) \times 10^3$ $^{40}\mathrm{K}$ atoms in a crossed dipole trap at the initial temperature $T/T_F=0.15(1)$, where $T_F$ is the Fermi temperature. The gas consists of an equal mixture of two spin components corresponding to the states $\ket {\uparrow} = \ket{m_F = -7/2}$ and $\ket{\downarrow} = \ket{m_F = -9/2}$ in the $F=9/2$ hyperfine ground-state manifold. Our sequence begins with loading the atoms into a blue-detuned three-dimensional optical lattice  with wavelength $\lambda_x=532\,$nm and lattice
constant $d =\lambda_x/2$  along the $x$ direction and $\lambda_\perp = 738 \,$nm in the transverse directions.
While the main lattice along $x$ is initially loaded to $20 \, E_{rx}$,   the transverse lattices are simultaneously ramped to a depth of $33 \, E_{r\perp}$, where they remain during the whole sequence to realize individual 1D systems. 
Here, $E_{rj}=\hbar^2 k_{j}^2/(2m)$ are the respective recoil energies with $j\in\{x,\perp \}$, $k_{j}= 2 \pi/\lambda_{j}$ denotes the corresponding wave vector and $m$ is the mass of $^{40}\mathrm{K}$. Holding the atoms in the deep initial lattice for $\SI{20}{\milli\second}$ dephases remaining correlations between neighboring sites, such that the resulting state can be approximated as a product state 
$|\psi_0 \rangle = \prod_{i\in {\rm trap}} \left(\hat{c}_{i\uparrow}^{\dagger}\right)^{n_{i\uparrow}} \left(\hat{c}_{i\downarrow}^{\dagger}\right)^{n_{i\downarrow}} |0\rangle$, where $\hat{c}_{i \sigma}^{\dagger}$ is the fermionic creation operator, $n_{i\sigma} \in \{0,1\}$, $\sigma \in \{ \uparrow, \downarrow \} $ and $i$ is the lattice-site index. The spin orientations are expected to be distributed randomly among the sites and the average number of atoms per lattice site in the center of the cloud $\langle \hat{n}_i \rangle= \sum_{\sigma} \langle \hat{n}_{i \sigma} \rangle$ is estimated to be $\langle \hat{n}_i \rangle \lesssim 0.9$~\cite{suppmat}, $ \hat{n}_{i\sigma}= \hat{c}_{i\sigma}^{\dagger}\hat{c}_{i\sigma}$ is the density operator. The fraction of atoms on doubly-occupied sites $n_d=N_{d}/(N_{s}+N_{d})$ in the initial state can be tuned via the interaction strength during the loading process employing a Feshbach resonance at $202.1\,\mathrm{G}$~\cite{suppmat}. Here, $N_{s}$($N_d$) denotes the number of particles on singly(doubly)-occupied sites. The dynamics starts with suddenly quenching the main lattice to $8 \, E_{rx}$. Simultaneously, the strength of the dipole trap is adjusted to compensate the anti-confining harmonic potential introduced by the lattice~\cite{suppmat}. Our system is then described by the homogeneous 1D FHM

\begin{equation}
\hat{H}=-J\sum_{i,\sigma=\uparrow,\downarrow}\left(\hat{c}_{i\sigma}^{\dagger}\hat{c}_{i+1\sigma}+ h.c. \right)+U\sum_{i}\hat{n}_{i\uparrow}\hat{n}_{i\downarrow} \,.
\end{equation}

\noindent After a variable expansion time $t$ the on-site population is frozen by suddenly increasing the lattice depth to $20 \, E_{rx}$. Subsequently, the cloud is imaged in-situ using high-field imaging either with or without removing doublons~\cite{suppmat}. By combining these images, the dynamics of singlons and doublons can be resolved individually.

{\it Quantum distillation.}
We characterize the dynamics by monitoring the singlon and doublon clouds as a function of the expansion time for an initial state with $n_d=0.40(2)$ [Figs.~\ref{fig:phase_separation}(a),\,(b)]. Isolated doublons are expected to become stable objects for interaction strengths that are large compared to the bandwidth $U\gg W$, $W=4J$~\cite{Rosch2008}, since in this case the interaction energy released in the doublon decay cannot be transferred into kinetic energy of singlons in low-order processes~\cite{Winkler2006,Rosch2008}. This is in agreement with our observations [insets in Figs.~\ref{fig:phase_separation}(a),\,(b)], where for $U=5J$ we witness a fast doublon decay of about $25\%$ in the early stages of the expansion $t\lesssim\!~\!5\tau$, which is accompanied by a compatible increase of the singlon number. In contrast, both numbers remain approximately constant for $U=20J$. Except for a small residual decay, which is attributed to light-assisted losses of doublons~\cite{Weiss1999}, this enables us to probe the dynamical phase separation of singlons and doublons at approximately constant doublon numbers.

We study the phase separation by extracting the cloud sizes $R_{s,d}(t)$ at half-width-at-half maximum (HWHM). We observe a rapidly-expanding singlon cloud, which has approximately doubled in size at $t\!=\!40 \tau$. In contrast, the doublon cloud size grows much slower and we even observe a weak shrinking of the cloud for $U\!=\!20J$. For comparison, we show the expected expansion of a fictitious cloud of non-interacting doublons expanding according to $J_{\rm{eff}}$~\cite{Rausch2017}. The difference highlights the non-trivial nature of this transient dynamics. The dynamical phase separation is even more evident in the comparison of the integrated line densities of singlons and doublons at $t\!=\!0$ and $t\!=\!40 \tau$ for our strongest interactions [Fig.~\ref{fig:phase_separation}(c)]. Clearly, the singlons expand significantly, while the doublons essentially remain in the center of the cloud. As a consequence, the ratio of atom numbers on doubly- and singly-occupied sites $N_d^c/N_s^c$ in the center of the cloud increases by about 40\% [Fig.~\ref{fig:phase_separation}(d)]. While this signal could in principle be caused by 1D systems with a low doublon fraction, a quantitative analysis based on our measured initial density distributions shows that their contribution to the signal is negligible~\cite{suppmat}. Hence, our data establish clear evidence for fermionic quantum distillation in the weak regime in a non-equilibrium mass-transport experiment.

\begin{figure}[t!]
	\includegraphics{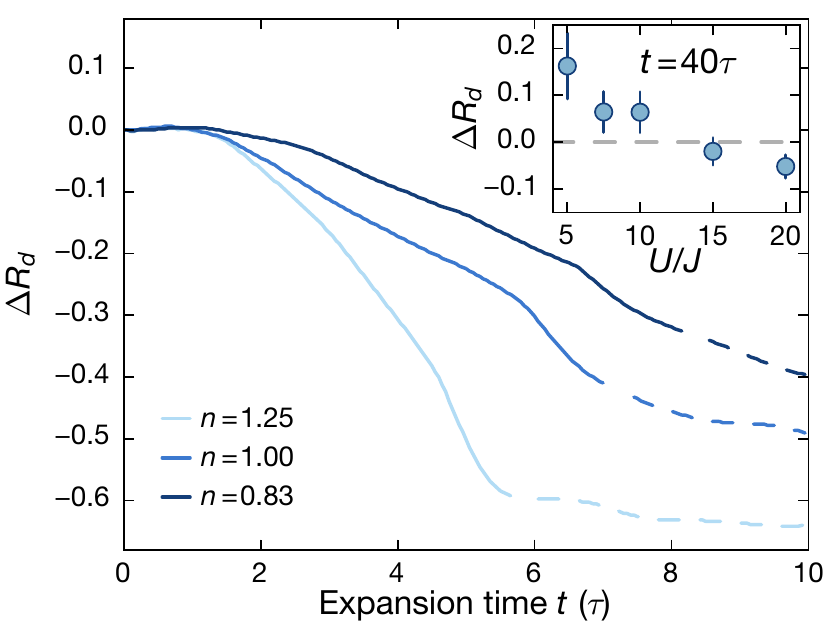}
	\caption{
{\bf tDMRG results for the relative doublon cloud size.} Main panel: Relative doublon could size $\Delta R_d(t)$ for three different initial uniform densities $n$ at $U/J=20$. The initial state consists of 12 singlons, 4 doublons and $\{0,4,8\}$ holons, respectively~\cite{suppmat}. The solid lines end at time $t_{\rm max}$, when the width of the singlon cloud increased to $\Delta R_s=0.8$. This value corresponds to the experimental one at $t\!=\!40\tau$.
Inset: Experimental data for $\Delta R_d$ as a function of the interaction strength at $t=40\tau$, which was evaluated using linear fits to the time traces $R_d(t)$ as shown in Figs.~\ref{fig:phase_separation}(a),\,(b)~\cite{suppmat}.}	\label{fig:atoms_core}
\end{figure}

{\it tDMRG results for transient dynamics.} Quantum distillation in the strong regime can further lead to a shrinking of the doublon cloud. The precise amount depends on the number of singlons initially confined in the doublon cloud, the initial density and the cloud size~\cite{Heidrich-Meisner2009,Herbrych2017,suppmat}. Here, we focus on the role of the initial density, which has the largest influence. Figure~\ref{fig:atoms_core} shows tDMRG simulations of the relative change of the doublon cloud size $\Delta R_d (t)=R_d(t)/R_d(0)-1$ as a function of time for different average initial densities $n=(N_s+N_d)/L_{\mathrm{init}}$ in an ideal box trap of length $L_{\mathrm{init}}$ for constant $n_d=0.4$~\cite{suppmat}. Negative values of $\Delta R_d$ indicate a shrinking of the doublon cloud, while $\Delta R_d > 0$ corresponds to an expanding doublon cloud. For the initial state with the largest density $n=1.25$, we observe a large decrease of $\Delta R_d(t)$. This effect is substantially reduced for smaller densities (Fig.~\ref{fig:atoms_core}) due to the presence of holons (empty sites), which remain trapped between doublons on the time scales of the quantum distillation process~\cite{Herbrych2017}. Additionally, the dynamics becomes slower, both due to holons and due to the larger cloud sizes used for simulations with smaller average densities~\cite{Herbrych2017}. Despite these quantitative differences, however, we find that the fundamental aspect of quantum distillation, i.e., the dynamical phase separation of singlons and doublons, is generally robust.

For comparison, we show the experimentally measured relative changes $\Delta R_d$ as a function of interaction strength [inset in Fig.~\ref{fig:atoms_core}]. For all interactions the time traces of the doublon HWHM are fitted with a linear function to calculate $\Delta R_d$ at the maximum expansion time $t=40 \tau$~\cite{suppmat}. We observe that $\Delta R_d(40\tau)$ approaches zero with increasing interaction strength and becomes slightly negative at $U/J=20$. In order to facilitate a comparison between experiment and numerics, where much smaller particle numbers are used, we define a time $t_{\rm max}$ for the simulation at which the relative singlon cloud size $\Delta R_s$ has reached the same value as in the experiment (Fig.~\ref{fig:atoms_core}). Our numerical results indicate that the contraction is not completed at this time. In the experiment this time is limited by the degree of flatness of the homogeneous potential. The remaining difference between the numerical and experimental results is most likely due to other initial-state properties, such as inhomogeneous density distributions and the averaging over several 1D systems with different initial-state properties~\cite{suppmat}.

\begin{figure}
	\includegraphics{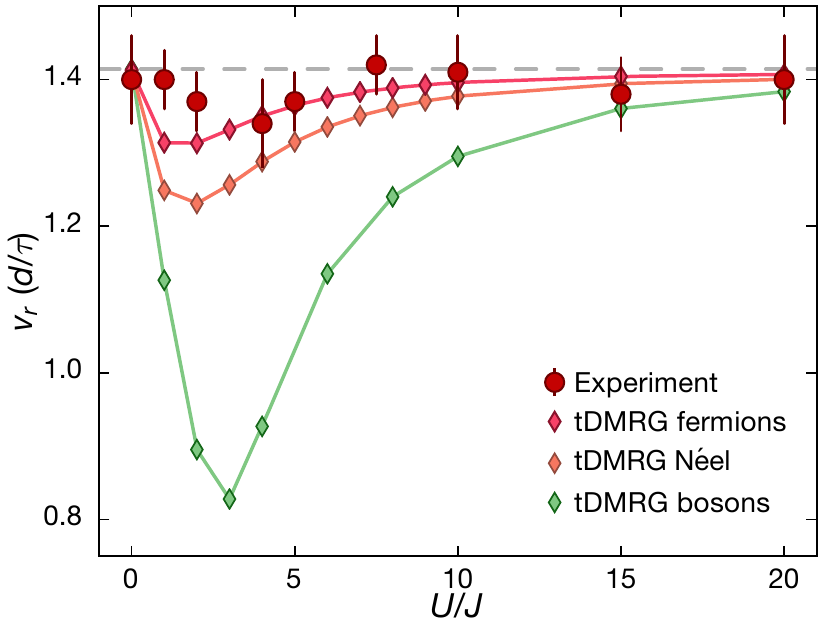}
\caption{\textbf{Radial expansion velocities \boldmath$v_r$.} Experiment (circles) and tDMRG simulations for fermions (red dark-shaded diamonds) and bosons (green light-shaded diamonds, from~\cite{Ronzheimer2013})
as a function of $U/J$. Solid lines are guides to the eye. The grey dashed line indicates $v_r=\sqrt{2} \, d/\tau$ in the limiting cases $U/J=0$ and $U/J \rightarrow \infty$. All initial states in the numerical simulations have a uniform average density of $n=1$ in a box with initial size $L_{\mathrm{init}}=10$.
}\label{fig:velos_exp_theory}
\end{figure}

{ \it Asymptotic dynamics and interaction effects.}
Here, we focus on the dynamics of the whole cloud for initial states with a negligible doublon fraction ($n_d<0.05$). We extract the second moment $r^2\!=\! \sum_{i} \langle \hat n_i\rangle (i_0 - i)^2 d^2/(N_s+N_d)$ of the time-dependent density distribution (see~\cite{suppmat} for details on the analysis), which is routinely computed in numerical simulations~\cite{Langer2012,Vidmar2013,Mei2016}; here $i_0$ is the center-of-mass of the initially trapped gas. From the time dependence of $r^2$, we extract the asymptotic radial velocity $v_r$ by fitting $\sqrt{r^2} =\sqrt{r_0^2+v_r^2t^2}$, where $r_0$ is the initial size of the cloud~\cite{suppmat}. Figure~\ref{fig:velos_exp_theory} shows $v_r$ as a function of $U/J$. We find $v_r\!=\!1.40(6) \, d/\tau$  for $U=0$ and $U=20J$, whereas for intermediate interactions $U \sim 3J$, the radial velocity decreases weakly. Note that for $U\gg W$, the mass transport in the 1D~FHM in the absence of doublons becomes identical to a non-interacting gas of spinless fermions and thus it behaves exactly like hardcore bosons in 1D with $v_r=\sqrt{2} \, d/\tau$~\cite{Ronzheimer2013}. The values in the limiting cases agree with these theoretical predictions for free fermions expanding from our initial state. Remarkably, compared to the Bose-Hubbard model~\cite{Ronzheimer2013}, the reduction of $v_r$ at intermediate interaction strengths is much weaker (Fig.~\ref{fig:velos_exp_theory}).

Starting from the limit of very strong interactions, the interaction dependence of $v_r$ can be understood in a two-component picture of independent singlon and doublon gases~\cite{Kajala2011,Sorg2014}: The dynamically generated doublons undergo a quantum distillation mechanism and are then inert on the time scales of the experiment. Thus, the more doublons are generated, the less kinetic energy is available for the rapidly expanding singlons. Focusing on the quantitative difference between the $v_r(U)$ curves for bosons and fermions, which is the main result of the data presented in Fig.~\ref{fig:velos_exp_theory}, two aspects are important. First, in the case of fermions, doublons can only be generated between sites with fermions of {\it different} spin orientation~\cite{Vidmar2013}. The initial state that has the most of such $\uparrow$-$\downarrow$ neighbors is the N{\'e}el state, and this initial state leads to the most pronounced minimum of $v_r$ (Fig.~\ref{fig:velos_exp_theory},~\cite{Vidmar2013}). In order to compare to the experiment, we average over many 1D systems with random spin orientations for a balanced spin mixture (dark red diamonds in Fig.~\ref{fig:velos_exp_theory}). This averaging leads to a weaker minimum in $v_r$ than for the N{\'e}el state and is in agreement with our experimental data. The second reason for the stronger minimum in $v_r$ for bosons is the fact that the interaction energy can become much larger, since larger local occupancies are possible~\cite{Ronzheimer2013}. In order to test whether the observed $v_r$ can primarily be understood as a function of the interaction energy in the system after the formation of doublons, we show data for different $U/J$ and different spin configurations versus interaction energy in the Supplemental Material (Fig.~\ref{fig:vr_vs_Eint} in~\cite{suppmat}). The data for bosons lie well outside the accessible range of interaction energies for fermions because of higher site occupations, but fall onto an extrapolation of the fermionic data. Hence, the integrability of the 1D FHM does not seem to be the dominant reason for the differences to the bosonic case. An interesting extension would be the calculation of expansion velocities by exploiting the integrability along the lines of~\cite{Mei2016,Bolech2012}, which we leave for future work.

{\it Summary and Outlook.} We investigated the sudden expansion of an interacting cloud of fermions. Starting from an initial product state with an appreciable doublon fraction, we observed a dynamical phase separation between singlons and doublons, theoretically known as fermionic quantum distillation in the weak regime. Additionally, we analyzed radial velocities for different interaction strengths using initial states consisting purely of singlons. We found a decrease of the radial velocities at weak interactions and attributed this effect to dynamically generated doublons. The weak decrease of radial velocities of expanding fermions compared to bosons is due to the Pauli principle leading to a crucial dependence of the radial velocities on the initial spin configuration. Future experiments could use the singlon and doublon resolved scheme to detect signatures of FFLO states~\cite{Kajala_FFLO11, Lu2012,Bolech2012} in the expansion velocity of the unpaired spin component. Moreover, it would be intriguing to observe the strong version of quantum distillation, resulting in the dynamical formation of low-entropy regions. This could be achieved by optimizing the initial-state properties and improved imaging techniques, such as microwave dressing to isolate central 1D systems, where the conditions for strong quantum distillation are best, or using quantum-gas microscopes~\cite{Cheuk2015, Haller2015}.

{\it Acknowledgments.}
 We acknowledge financial support by the European Commission (UQUAM grant no. 319278, AQuS) and the Nanosystems Initiative Munich (NIM) grant no. EXC4. J.S. and F.H.-M. were supported by the DFG (Deutsche Forschungsgemeinschaft)
Research Unit FOR 1807 under grant no. HE 5242/3-2 and by startup funds via SFB 1073 at the University of G\"ottingen. J.H. was supported by the US Department of Energy (DOE), Office of Basic Energy Sciences (BES), Materials Sciences
and Engineering Division. 



%


\cleardoublepage

\setcounter{figure}{0}
\setcounter{equation}{0}
\setcounter{page}{1}

\renewcommand{\thepage}{S\arabic{page}} 
\renewcommand{\thesection}{S\arabic{section}} 
\renewcommand{\thetable}{S\arabic{table}}  
\renewcommand{\thefigure}{S\arabic{figure}} 
\renewcommand{\theequation}{S\arabic{equation}} 

\renewcommand{\thesection}{}
\renewcommand{\thesubsection}{S\arabic{subsection}}

\section*{\large{Supplemental Material}}
\setlength{\intextsep}{0.8cm} 
\setlength{\textfloatsep}{0.8cm}

\newcommand{\jan}[1]{{\color{cyan} [Jan S.: #1]}}

\subsection{Initial state preparation}

\subsubsection{Experimental sequence}
The initial state preparation for the measurements in the main paper starts with loading the atoms into the three-dimensional (3D) optical lattice using a sequence of different linear ramps. First, within \SI{7}{\milli\second}, the lattice along the $x$~direction and the transversal lattices are ramped to a depth of $1 \,E_{rx}$ and $1 \,E_{r\perp}$, respectively. Then, after waiting for \SI{100}{\milli\second}, the depth of all three lattices is further increased to $8 \,E_{rx}$ and $8 \, E_{r\perp}$, respectively, during \SI{75}{\milli\second}. Finally, the lattice along the $x$~direction is ramped up to $20 \, E_{rx}$ and the transversal lattices reach their final depth of $33 \, E_{r\perp}$ within  \SI{15}{\milli\second}, freezing the populations of singlons and doublons. Afterwards, an additional superlattice along the $x$~direction is added at a depth of  $20 \, E_{s}$, where $E_s=h^2 /(2m \lambda_{s}^2)$ is the recoil energy of the superlattice with wavelength $\lambda_{s}=1064\,\mathrm{nm}$. The phase of the superlattice was set so as to create tilted double wells along the $x$~direction, in order to decrease residual dynamics and remaining correlations between neighboring sites. 

While holding the atoms in the deep 3D optical lattice for \SI{25}{\milli\second}, both the dipole trap strength and the magnetic field strength are adjusted to their target values during the expansion of the cloud. We ramp the magnetic field within \SI{15}{\milli\second} to change the scattering length from $a_s=-20 a_0$ (attractive loading to generate initial states with doublons) or $a_s=140 a_0$ (repulsive loading to realize initial states without doublons) to the scattering length which sets the desired interaction strength during the expansion of the cloud in the lattice (see Sec.~\ref{sec:control}). Moreover, the dipole traps along the $x$ and the $y$~direction [trap frequency of $\omega_x=\omega_y=\omega= 2 \pi \times \SI{54(1)}{\hertz}$, measured in the $(8 E_{rx},8 E_{r\perp}, 8 E_{r\perp})$ deep lattice during loading] are ramped to zero within \SI{22}{\milli\second}, whereas the dipole trap along the $z$~direction [trap frequency of $\omega_z= 2 \pi \times \SI{184(2)}{\hertz}$, measured in the $(8 E_{rx},8 E_{r\perp}, 8 E_{r\perp})$ deep lattice during loading] is ramped within \SI{22}{\milli\second} such that the lattice potential is flat during the expansion (see Sec.~\ref{sec:flat}). The preparation sequence terminates by removing the superlattice and quenching the lattice along the $x$~direction within \SI{10}{\micro \second} from $20 \, E_{rx}$ to $8 \, E_{rx}$, in this way initiating the expansion of the cloud. 

\subsubsection{Characterization}

We characterize the initial state by estimating the renormalized cloud size $R_{sc} = R/(\gamma_y \gamma_z N_\sigma)^{1/3}$  and the dimensionless compression $E_t/(12J)$ during loading, where $E_t=V_t [3 \gamma_y \gamma_z N_\sigma / (4 \pi )]^{2/3}$ and $V_t=  m \omega^2 d^2/2$. Here, $\gamma_y=738/532$ takes into account the different lattice constants between the $x$~direction and the transversal $y$ and $z$~directions, $\gamma_z=\omega_z/ \omega=184/54$ takes into account the different harmonic confinement along the $z$~direction compared to $x$ and $y$~directions, $N_\sigma = 15(1) \times 10^3$ is the number of atoms per spin state $\sigma$, $m$ is the mass of $^{40}K$ and $d$ is the lattice constant along the $x$~direction. The parameters $R_{sc}$ and $E_t$ were previously used to distinguish between the Mott-insulating and the metallic regime of the Fermi-Hubbard model for repulsively-interacting fermions~\cite{Schneider2008}. The renormalized cloud size is a rough estimate for the distance between two particles in the same spin state and only depends on the dimensionless compression, the interaction strength and the entropy set by the temperature in the pure harmonic trap. The dimensionless compression can be understood as the ratio between the characteristic trap energy $E_t$, which is the Fermi energy of a non-interacting gas in the zero-tunneling limit, and the bandwidth $12J$ in a 3D optical lattice. We estimate  $R_{sc}=0.9(1) d$ and $E_t/(12 J)=0.1(1)$ in the $(8 E_{rx},8 E_{r\perp}, 8 E_{r\perp})$ deep lattice. This means that we work in the metallic regime with a mean density in the center of the cloud of $\langle \hat{n}_i \rangle <1$.

\subsection{Control of the doublon fraction}
\label{sec:control}

\begin{figure}[t!]
	\includegraphics{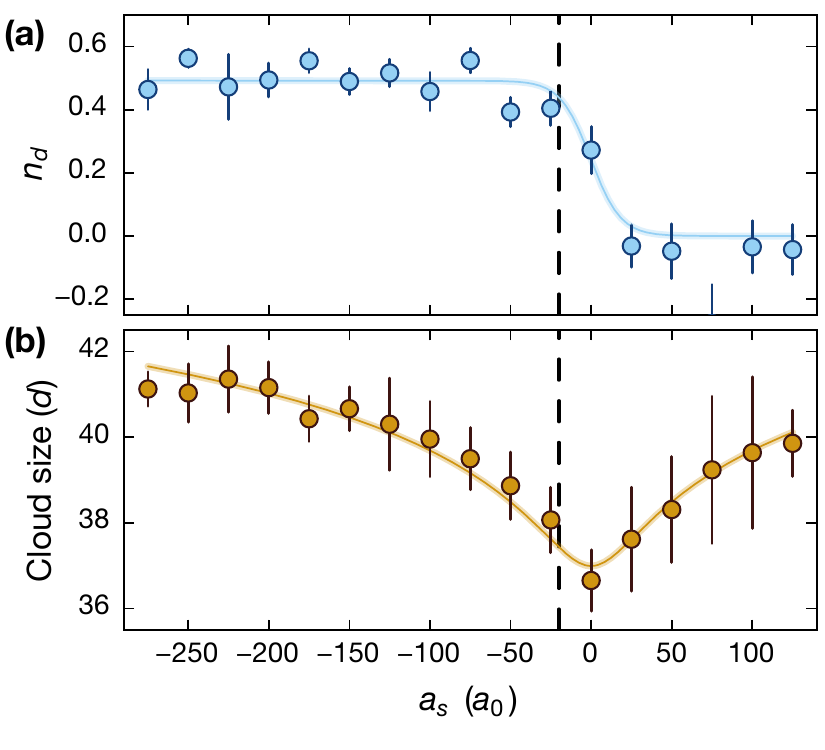}
	\caption{\textbf{Control of the doublon fraction:} (a) Fraction of atoms on doubly-occupied sites $n_d$ depending on the loading scattering length in units of the Bohr radius $a_0$ and (b) the resulting cloud size in units of the lattice constant $d$. To prepare initial states with a large number of doublons, we use a loading scattering length of $a_s=-20 \, a_0$ (dashed vertical line), which yields a significant fraction of atoms on doubly-occupied sites $n_d=0.40(2)$ and a minimal cloud size. Error bars denote the standard-error-of-the-mean.}
	\label{fig:doublons_ms}
\end{figure}

The doublon fraction of the initial state can be set by properly adjusting the scattering length $a_s$ during the loading of the atoms into the 3D optical lattice. While large repulsive interactions ($a_s > 0$) result in initial states with a negligible doublon fraction, attractive interactions ($a_s < 0$) favor initial states with a discernible doublon fraction, as illustrated in Fig.~\ref{fig:doublons_ms}(a). Creating initial states with a large doublon fraction, however, also increases the number of holes, since a pair of neighboring singlons is converted into a doublon and a hole. In order to generate initial states with an appreciable fraction of atoms on doubly-occupied sites [$n_d=0.40(2)$], we prepare the initial state with $a_s= -20a_0$. This value was chosen in order to maximize the doublon fraction, while at the same time minimizing the cloud radius and therefore maximizing the density. As shown in Fig.~\ref{fig:doublons_ms}(b), the cloud radius is the same as the one, where the scattering length was set to $25 a_0$ during loading; with $U/(12J)=0.12$ in the $(8 E_{rx},8 E_{r\perp}, 8 E_{r\perp})$ deep lattice. The measured cloud radius [Fig.~\ref{fig:doublons_ms}(b)] is in agreement with the values obtained in Ref.~\cite{Schneider2008}, where, for our parameters, the central density of the cloud was shown to be $\langle \hat{n}_i \rangle \lesssim 0.9$. For initial states without doublons ($n_d < 0.05$), we use a loading scattering length of $a_s= 140a_0$.

\subsection{Creating a flat potential for the expansion}

\label{sec:flat}
At the end of the initial state preparation, the atomic cloud is confined by three dipole traps: one along each of the horizontal directions, $x$ and $y$ ($x$ is the longitudinal direction of the tubes), and one along the vertical $z$~direction. The vertical dipole trap has a Gaussian beam waist of $150 \, \mu\mathrm{m}$. The horizontal dipole traps are elliptical with waists of $30 \, \mu \mathrm{m}$ in the vertical and waists of $300 \, \mu \mathrm{m}$ in the horizontal direction. The optical lattices along all spatial axes have beam waists of $150 \, \mu \mathrm{m}$ (as the vertical dipole trap) and are blue detuned, providing an anti-confining potential.  A flat potential along the $x$~direction during the expansion of the cloud can be generated by choosing the strength of the vertical dipole trap such that it compensates the anti-confinement of the optical lattices. As the horizontal dipole traps have different beam geometries, they cannot be used to compensate the anti-confinement, and therefore, they are switched off during the expansion measurements. Creating a flat potential requires optimizing both the $z$ dipole beam alignment and its strength to maximize the in-situ cloud size after a long evolution time (see also Ref.~\cite{Ronzheimer2013}). However, a completely flat potential cannot be implemented using this method, as the harmonic contributions of the overall potential can only be canceled within a certain area.

\subsection{Singlon- and doublon-resolved analysis}

\begin{figure}
	\includegraphics{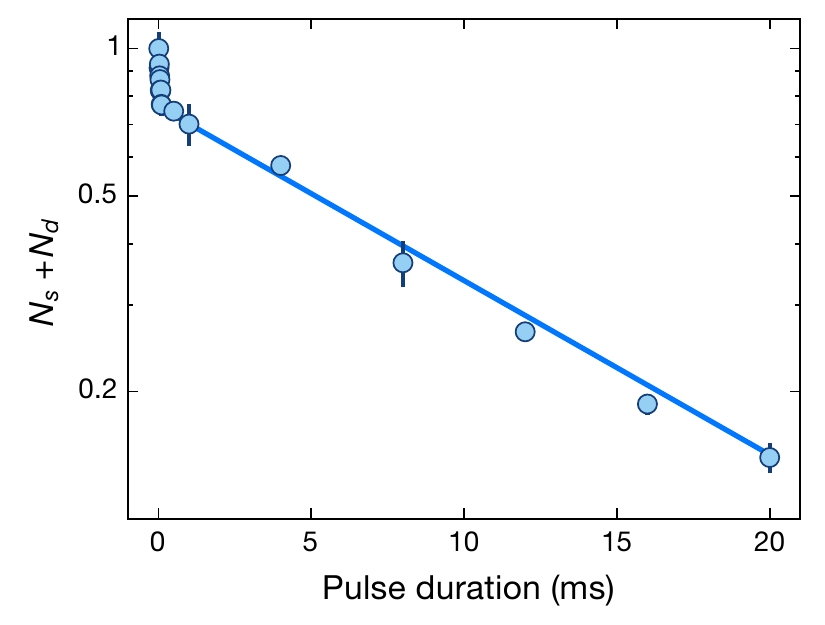}
	\caption{\textbf{Atom number decay in the presence of an additional near-resonant laser pulse:} Total atom number $N_s + N_d$ as a function of the pulse duration. The data was taken for $U=20J$ after an expansion time in the lattice of $t=40 \tau$ at a detuning of the pulse of $\Delta=\SI{364}{\mega\hertz}$. Solid lines show the fit using the sum of two exponentials.}
	\label{fig:BlastDur}
\end{figure}

Doublons can be removed from the lattice using an additional laser pulse, which is blue-detuned by an amount $\Delta$ from the imaging transition of $^{40}$K ($|F=9/2,m_F=-9/2\rangle \rightarrow |F'=11/2,m_{F'}=-11/2\rangle$). As a result, atoms on doubly-occupied sites are lost due to light-assisted collisions. As indicated by the semi-log-plot in Fig.~\ref{fig:BlastDur}, the total atom number shows a bimodal decay with well-separated time scales, which we characterize by fitting a sum of two exponential functions to the atom number $N_s (t)+ N_d (t) = N_s(t=0)\exp\left(-t/\tau_s \right) + N_d(t=0)\exp\left(-t/\tau_d \right)$. We extract a fast decay with a lifetime of $ \tau_d = \SI{40(10)}{\micro\second}$, which is attributed to the loss of atoms on doubly-occupied sites due to light-assisted collisions and an additional slow decay  with a lifetime of $  \tau_s=\SI{12(1)}{\milli\second}$, which results from the loss of atoms on singly-occupied sites due to off-resonant photon scattering. For all experiments, the duration of the light pulse was set to $\SI{150}{\micro\second}$. The detuning $\Delta$ of the pulse with respect to resonant imaging light depends on the magnetic field used to set the final interaction strength with the Feshbach resonance and varies between $\Delta (U\!=\!5J)=\SI{296}{\mega\hertz}$ and $\Delta (U\!=\!20J)=\SI{364}{\mega\hertz}$. We found that the different detunings have only a negligible effect on the doublon lifetimes. By subtracting the optical density of successive pictures with and without light pulse, doublon- and singlon-resolved dynamics can be analyzed individually.

\subsection{Estimation of tubes with few doublons}

Absorption imaging along the $z$~direction results in averaging over many individual realizations of 1D systems with different density distributions. Therefore, the increase of $N_d^c/N_s^c$ in Fig.~\ref{fig:phase_separation}(d) of the main text could in principle be caused by 1D systems that contain singlons but only a very small amount of doublons. In these systems, the expansion would decrease the singlon number in the center $N_s^c$, without mediating doublon dynamics. In order to estimate the contribution of 1D systems with negligible doublon fraction to the ratio $N_d^c/N_s^c$, we can approximate the initial shape of the cloud in the lattice with an ellipsoid, where the ratio of the principal axes is set by the trap frequencies. Using an Abel transform, the full 3D density distribution of singlons and doublons can be reconstructed. The width of the singlon and doublon distribution along the $x$~direction thereby translates into a width along the $z$~direction by implying the symmetry of the ellipsoid. As a result, the extent of the cloud along the $z$~direction amounts to about 30 individual planes, where only the four outermost planes are primarily occupied with singlons. Assuming that the dynamics and therefore the change in the ratio $N_d^c/N_s^c$ is solely governed by singlons expanding in these outermost 1D systems we obtain a conservative upper bound of $N_d^c/N_s^c \lesssim 10\%$. The observed effect reported in the main text [Fig.~\ref{fig:phase_separation}(d)] is much larger than this upper bound and therefore clearly indicates a dynamical phase separation in 1D systems with singlons and doublons.

\subsection{Fits to the time traces of $R_d$}

\begin{figure}[t]
	\includegraphics[width=1.0\linewidth]{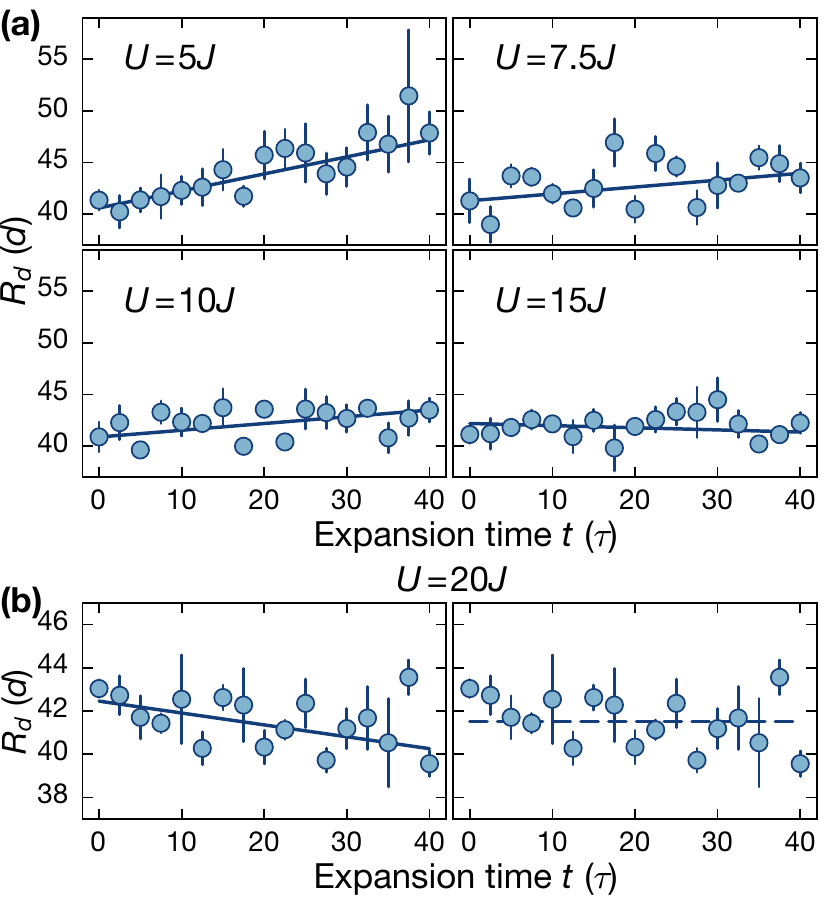}
	\caption{\textbf{Time traces of $R_d$ with linear fits}: (a) Time traces of the HWHM of the doublon cloud $R_d$ for different interactions. Solid lines indicate a linear fit to the data. (b) Time trace of $R_d$ for $U=20J$ with a linear fit resulting in $\chi_{\mathrm{red}}^2=0.61$ (left) and a fit of a constant function (dashed line) resulting in $\chi_{\mathrm{red}}^2=0.84$ (right).  }
	\label{fig:FitsFig3}
\end{figure}

In order to quantify the dynamics of the width of the doublon cloud in Fig.~\ref{fig:phase_separation}, we fit the time traces of $R_d$ with a linear function $f(t)=at+b$. The fits to the data are shown in Fig.~\ref{fig:FitsFig3}(a) for increasing interactions and additionally in the left panel of Fig.~\ref{fig:FitsFig3}(b) for the strongest interaction of $U=20J$. While the fits indicate an increase of $R_d$ for weak interactions of $U=5J$, stronger interactions result in a slower spreading of the doublon cloud and for $U=20J$, the fit clearly indicates a contraction of the doublon cloud as shown in the  left panel of Fig.~\ref{fig:FitsFig3}(b). In order to test the goodness of the linear fit to the doublon cloud size for strong interactions of $U=20J$, we compare it to fitting a constant function to the doublon cloud size in the right panel of Fig.~\ref{fig:FitsFig3}(b). A common quantitative estimate for the goodness of a fit is $\chi_{\mathrm{red}}^2$, which is defined as 

\begin{equation}
\chi_{\mathrm{red}}^2=\nu^{-1} \sum_{i=1}^{N} \left(\frac{R_d(t_i) -g(t_i)}{\sigma_i} \right) ^2 \, ,
\end{equation}

\noindent where $R_d(t_i)$ is the doublon cloud size at time $t_i$, calculated from the mean of four data points, $\sigma_i$ is the corresponding standard deviation for each $R_d(t_i)$, $g(t_i)$ is the value of the fit function at time $t_i$ and $\nu$ is the difference between the total number $N$ of discrete points $t_i$ ($N=17$ for all time traces of the doublon cloud size) and the number of fitting parameters. For the linear fit, we obtain $\chi_{\mathrm{red}}^2=0.61$, whereas the fit with the constant function yields a slightly larger  $\chi_{\mathrm{red}}^2=0.84$. This $\chi_{\mathrm{red}}^2$ analysis indicates that the decreasing linear function describes the data better than a constant function, even after accounting for the increased number of fit parameters. Using the resulting fit parameters, we calculate the relative change in doublon cloud size $\Delta R_d= a/b \cdot 40 \tau $ for all interactions and show the results in the inset of Fig.~\ref{fig:atoms_core} in the main text. The error of the linear fitting parameters indicates one sigma confidence intervals and the errorbars of the relative doublon cloud size are calculated by using Gaussian error propagation.

\subsection{Extracting radial velocities}

\begin{figure}[t]
	\includegraphics{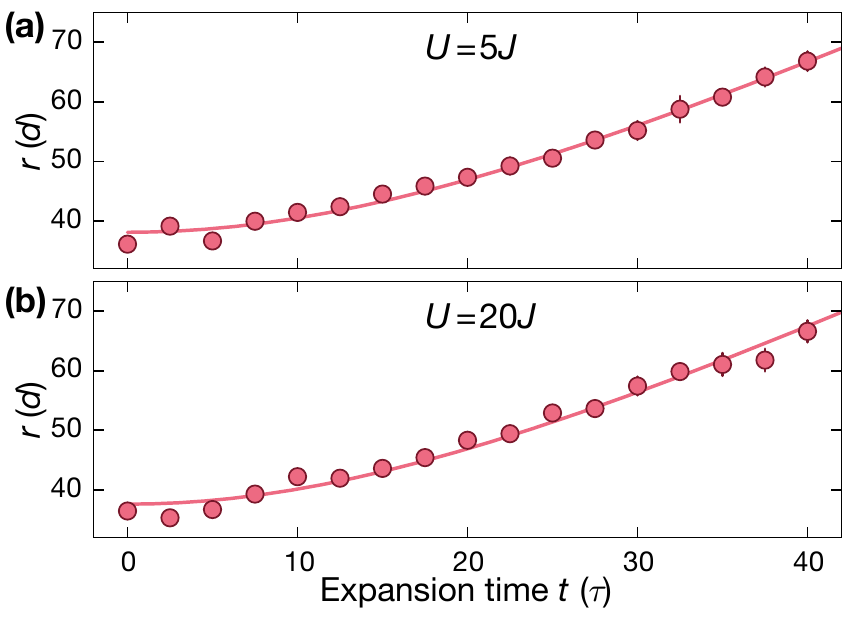}
	\caption{\textbf{Second moments $ r^2 $}: Exemplary time traces of the cloud size $r$ for (a) $U=5J$ and for (b) $U=20J$. Solid lines are fits to the time traces to extract the radial velocities.}
	\label{fig:radial_velo}
\end{figure}

The size of the cloud can be characterized by measuring the second moment $ r^2 $, which is defined as  $ r^2   = \sum_{l} \rho_{l} (l-l_{c})^{2} \cdot 8.3^2 d^2$, where $\rho_{l}$ is the normalized optical density at pixel $l$ with distance $|l-l_{c}|$ from the central pixel $l_{c}$ of the 1D integrated line densities and the factor $8.3$ converts pixels to lattice sites in units of the lattice spacing $d$. The second moment is less affected by details of the density distribution than $R_d$ and $R_s$, since it takes into account the density of  the whole cloud and not only the height of the cloud at a chosen width. This, however, leaves it more susceptible to noise, compromising its applicability in the analysis of the doublon clouds (Fig.~2 of the main text). The second moment $r^2 $ is extracted by subtracting the background from the raw integrated line densities and then summing over the integrated line densities of the cloud from the center outwards, until $ r^2 $ saturates. The cloud size at every time step is averaged over five measurements and error bars denote the standard error of the mean. Exemplary time traces of $r=\sqrt{ r^2 }$ are shown in Fig.~\ref{fig:radial_velo}(a) for $U=5J$ and in Fig.~\ref{fig:radial_velo}(b) for $U=20J$. From the time dependence of $ r^2 $, we extract the radial velocity $v_r$ by fitting $\sqrt{ r^2 } =\sqrt{r_0^2+v_r^2t^2}$, where $r_0$ is the second moment of the cloud at $t=0$. We do not use $\tilde{r}$ as defined in Eq.~\eqref{eq:tilde_r} in the following Sec. S8.3, since this would require a sufficiently accurate  knowledge of $r_0$. In the non-interacting case, the radial velocity of an initially localized particle yields $v_r=\sqrt{2} \, d/\tau$ (in agreement with our experimental results for $U=0$) and are an average in quasi-momentum space over the group velocities weighted with the momentum-distribution function~\cite{Langer2012,Vidmar2013}.

\subsection{tDMRG simulations}

\subsubsection{Method}

The non-equilibrium expansion is studied by means of the time-dependent density-matrix renormalization group (tDMRG) method 
(see Refs.~\cite{White2004,Daley2004,Vidal2004,Schollwock2011} for details) using 
a  Trotter-Suzuki decomposition scheme for the time propagation.
The results presented in Fig.~\ref{fig:atoms_core} of the main text were obtained with 
 a second-order Trotter-Suzuki decomposition, up to $M=6000$ DMRG states, a discarded weight of $10^{-6}$ and $\delta t =0.05 \tau$.
For the data shown in Figs.~\ref{fig:r_inst} and \ref{fig:vr_vs_Eint}, $\delta t =0.1 \tau$ was used.
We ensure the numerical accuracy of the results by varying the time step, the discarded weight during the time evolution and the maximum number of states. Some of our tDMRG implementations make use of the tensor library developed in \cite{Dorfner2016}.

The initial states are prepared on a lattice of $L$ sites as a product state, i.e., 
\begin{equation}
|\psi_{0}\rangle = \prod_{i=i_L,\dots,i_R}
\left(\hat{c}_{i\uparrow}^{\dagger} \right)^{n_{i\uparrow}} \left(\hat{c}_{i\downarrow}^{\dagger}\right)^{n_{i\downarrow}}|0 \rangle\,,
\label{initps}
\end{equation}
with $i_L=L/2-L_{\mathrm{init}}/2$, $i_R=L/2+L_{\mathrm{init}}/2$, where $L_{\mathrm{init}}$ is the size of the region with a nonzero density at $t=0$  and ${n}_{i\sigma}\in\{0,1\}$ are specially chosen integers. Thus, each site inside the initially confined region is occupied by exactly one singlon, doublon, or holon (empty site). 
The experimental results are averages over many 1D tubes in each measurement of the cloud and, as a consequence, various singlon/doublon/holon configurations are probed simultaneously. In order to account for this, the results presented in Fig.~\ref{fig:atoms_core} of the main text are averaged over $120$ simulations, each starting from different (random) distribution of particle configurations. The latter ensures an approximately uniform density in the initially ($t=0$) confined region. Note that the results contain an error due to the sampling  over a subset of all possible configurations. This small error is comparable in magnitude to our numerical accuracy at the maximal considered time (order of a few percent). In addition, each sample is prepared with a random distribution of spin-up and spin-down fermions under the constraint of $ N_{\downarrow} =  N_{\uparrow}$. Due to the entanglement increase in tDMRG simulations~\cite{Schollwock2011}, it is unfeasible to simulate the full distribution of 1D tubes realized in the experiment for realistic particle numbers and time scales.

\subsubsection{Initial states with doublons}

 We keep the total number of doublons fixed to $N_d/2= \sum_i\langle  \hat{n}_{i\uparrow} \hat{n}_{i\downarrow} \rangle =4$ and set the singlon-to-doublon ratio to $ 2N_{s}/ N_{d}=3$, where $ N_{s}= N-N_d$ with $N=\sum_{i\sigma} \langle \hat{n}_{i\sigma} \rangle$. Note that $N_d$ denotes the number of atoms on doubly-occupied sites, hence, the number of doublons is $N_d/2$. We work with $L=60$ sites and we only consider times before reflections off the boundaries occur.

\begin{figure}[t]
	\includegraphics{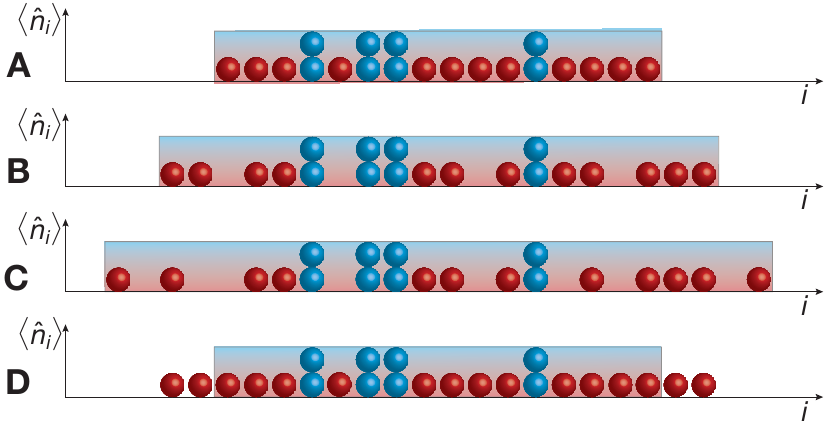}
	\caption{\textbf{Schematics of the different initial states with doublons}: {\bf A}: Box trap with singlons and doublons, 
{\bf B} and {\bf C}: Box trap containing doublons, singlons and holons,
{\bf D}: singlons and doublons as in case {\bf A} but surrounded by singlon wings. The shaded area depicts the  region 
		in which we randomly distribute doublons, singlons, or holons, keeping $N_d/2=4$ and $N_s=12$ fixed. Outside those regions, we only allow for singlons (case {\bf D}) or no atoms. 
	}\label{suppfig_schematic}
\end{figure}

As is clearly visible in Fig.~3 of the main text, the density of the initial state highly influences the  quantum-distillation dynamics. Here, we show how additional details of the initial cloud affect the achievable doublon contraction and the transient expansion dynamics.
In order to account for typical aspects of the experiment, we consider four types of initial states (see also Fig.~\ref{suppfig_schematic}):
\begin{itemize}
\item[{\bf A}:] A box trap with $12$ singlons and $4$ doublons and an initial box size of $L_{\mathrm{init}}=16$,
\item[{\bf B}:] A box trap with $12$ singlons, $4$ doublons, and $4$ holons and an initial box size of $L_{\mathrm{init}}=20$,
\item[{\bf C}:] A box trap with $12$ singlons, $4$ doublons, and $8$ holons and an initial box size of $L_{\mathrm{init}}=24$,
\item[{\bf D}:] case {\bf A} surrounded by singlon wings, i.e., two additional singlons placed on the left and the right side of the configurations used  in {\bf A},
\end{itemize}

Case {\bf A} corresponds to the box trap with solely singlons and doublons. This idealized setup was previously investigated in theoretical studies~\cite{Heidrich-Meisner2009, Herbrych2017}. 
Cases {\bf B} and {\bf C} also include 4 or 8 holons, respectively, and thus $L_{\rm init}$ increases as we go from  {\bf A} to {\bf C}. At the same time, the average density
goes down. We choose to keep $N_d/2$ fixed since we are interested in how close we can get to the maximum doublon contraction (i.e., all four doublons on neighboring sites).
These are the cases for which we show $\Delta R_d$ in Fig.~3 of the main text. The presence of holons can be viewed as imperfections resulting  during the loading process and initial-state preparation.  
Case {\bf D} accounts for the inhomogeneous shape of the initial fermionic cloud  resulting from the trapping potential. 

\begin{figure}[t]
	\includegraphics{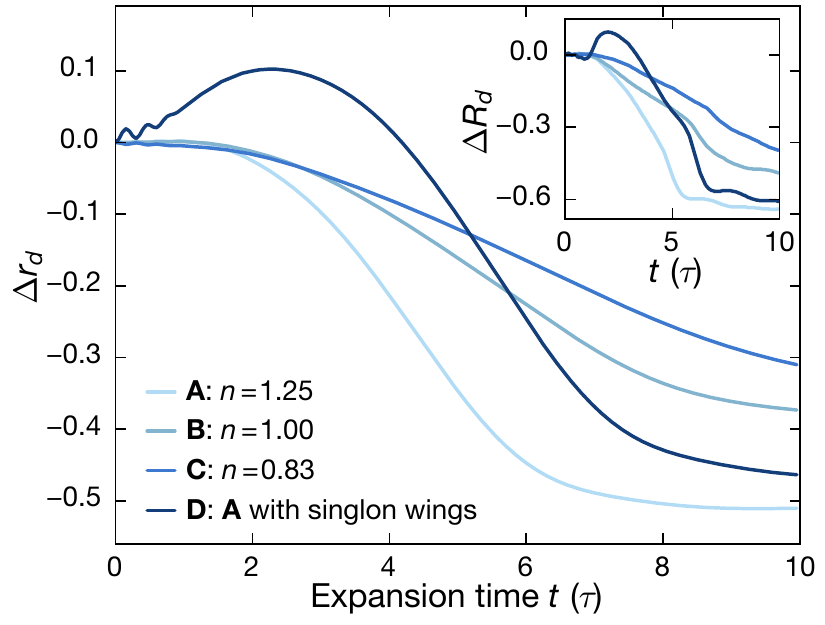}
	\caption{\textbf{Doublon cloud radius}:  Time evolution of  $\Delta {r}_d$ for $U=20J$ and various initial configurations. Inset: Time traces of $\Delta{R}_d$ (relative change in the HWHM). 
}\label{suppfig}
\end{figure}

Complementary to the main text, where we show the HWHM $R_d$, here, we discuss the doublon cloud radius defined as $r_d(t)=\sqrt{(1/N_d(t))\sum_i n^{d}_i(t)(i-i_0)^2}$, with $i_0$ the center of mass (here,  $i_0=L/2 + 0.5$) and $n_i^d=\langle  \hat{n}_{i\uparrow} \hat{n}_{i\downarrow}\rangle$ is averaged over all random configurations. In the main panel of Fig.~\ref{suppfig}, we present the relative change of the cloud radius $\Delta {r}_d(t)=r_d(t)/r_d(0)-1$ for an interaction strength of $U/J=20$. 
We stress the  main observations: (i) The time traces of  $\Delta {r}_d$ behave qualitatively very similar to the time traces of $\Delta{R}_d$ (relative change in the HWHM shown in the inset of Fig.~\ref{suppfig}). 
(ii) The presence of holons, cases {\bf B} and {\bf C}, reduces the achievable minimal cloud size on the time scales of the quantum distillation~\cite{Herbrych2017}.
(iii) The presence of additional singlons at the edges of the cloud (case {\bf D}) allows doublons to move outwards on the singlon wings in the early stage of the expansion. As a consequence, the presence of singlon wings delays the contraction of doublons. 

\subsubsection{Initial states without doublons}
\label{sec:wo_doublons}
The initial states without  doublons are prepared by enforcing $n_{i\uparrow}+n_{i\downarrow}=1 $ in Eq.~\eqref{initps} on a  lattice of $L=100$ sites with $N=10$ particles placed in the center of the system. Note that in this case, we keep the particle number fixed and do not consider holons in the initial state,  while our results are averaged over all possible spin configurations with $N_\uparrow = N_\downarrow$.

\begin{figure}[t!]
	\centering
	\includegraphics[width=1.0\linewidth]{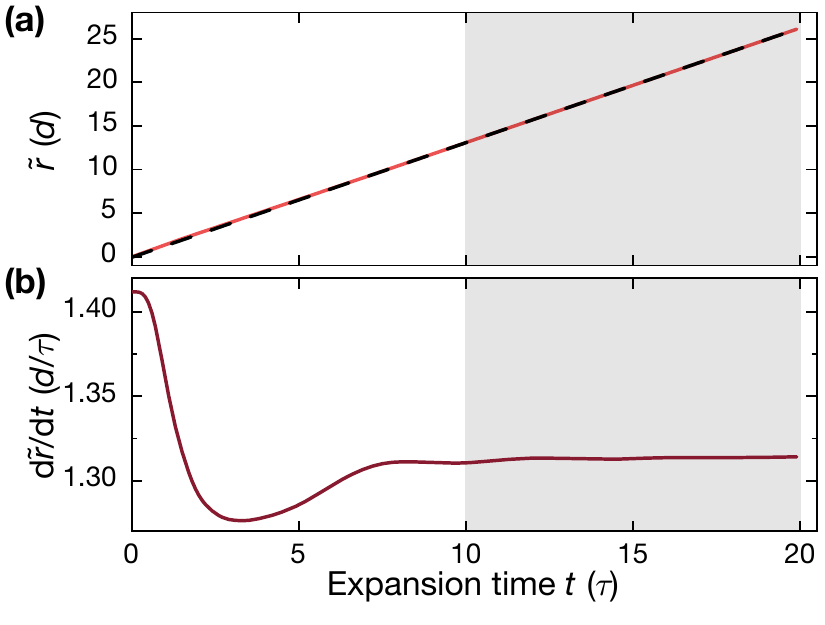}
	\caption{\textbf{Extraction of the radial velocity $v_r$ from the tDMRG data}: (a) $\tilde r$ as defined in Eq.~\eqref{eq:tilde_r} as a function of time indicated as a red solid line and linear fit to the curve indicated by a dashed black line. (b) Time derivative of $\tilde r$ to determine the time when the transient behavior is finished and $\tilde r$ is approximately linear in time. The shaded region represents the time interval where the linear fit is done.}
	\label{fig:r_inst} 
\end{figure}

In order to extract the velocities, we first calculate $\tilde r(t)$ via:
\begin{align}
  r^2(t) = \frac{1}{N} \sum_{i=1}^L \langle \hat n_i\rangle (i_0 - i)^2, \\
  \tilde r(t) = \sqrt{ r^2(t) - r^2(0)}, \label{eq:tilde_r}
\end{align}
where $\langle \hat n_i\rangle$ is the density at site $i$ averaged over all initial spin configurations.  The procedure to extract the asymptotic velocities is illustrated in Fig.~\ref{fig:r_inst} where the shaded region indicates the fitting window. 
Clearly, $\tilde r$ is essentially linear in time and in the fitting window the velocity has settled to a constant value, as shown in Fig.~\ref{fig:r_inst}(b). 

\begin{figure}[b!]
	\centering
	\includegraphics{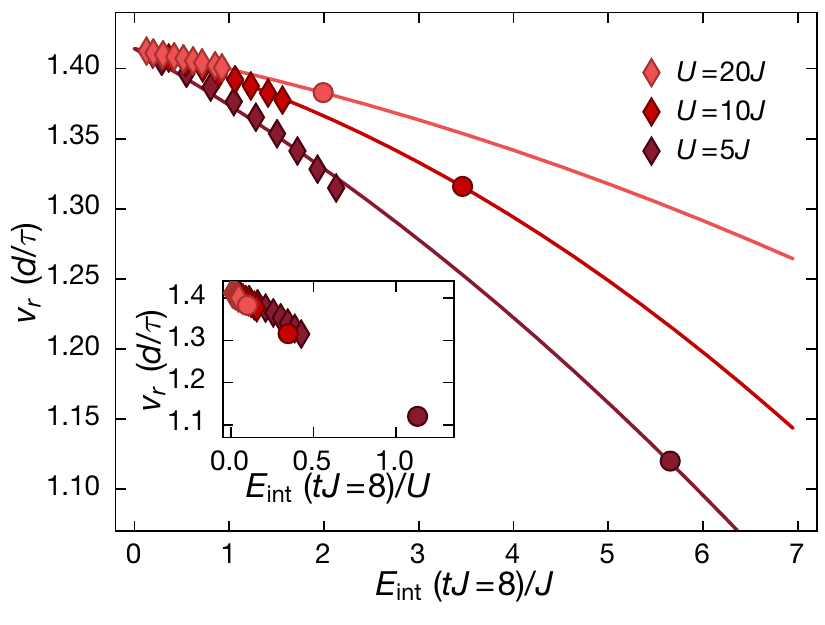}
	\caption{\textbf{Radial velocity as a function of the interaction energy for different interaction strengths}: The three colors correspond to interaction strengths $U/J = 5, 10, 20$. Diamonds correspond to fermions, circles correspond to bosons. Diamonds of the same color correspond to different numbers of domain walls in the initial state (only one domain wall to nine domain walls from left to right). Solid lines are quadratic fits to the data. The inset shows the same data but versus $E_{\mathrm{int}}/U$.}
	\label{fig:vr_vs_Eint}
\end{figure}

The different initial states can be discriminated by the number of domain walls that is the number of times an up-spin particle is next to a down-spin particle. For $N = 10$ particles in the initial state, the minimum number of domain walls is one and the maximum number is nine. We calculate the velocities for different numbers of domain walls in the initial state as before by first averaging over all configurations with the same number of domain walls and then  extract $v_r$. In Fig.~\ref{fig:vr_vs_Eint}, we plot the velocities for different numbers of domain walls versus the interaction energy at time $tJ = 8$ as diamonds (one to nine domain walls in the initial state from left to right)~\cite{Vidmar2013}. The interaction energy is defined as:
\begin{align}
E_{\rm{int}} = U \sum_{i=1}^L \langle \hat n_{i\uparrow} \hat n_{i\downarrow} \rangle.
\end{align}

We also carried out a simulation of the Bose-Hubbard model where we start with a region of 10 singly-occupied sites in the middle of an otherwise empty lattice~\cite{Ronzheimer2013,Vidmar2013}. For the Bose-Hubbard model, this is the unique product state with one boson per site
and doublons and higher site occupancies can be generated dynamically in every site.
 The interaction energy for the Bose-Hubbard model is defined as:
\begin{align}
  E_{\rm{int}} = \frac{U}{2} \sum_{i=1}^L \langle \hat n_i (\hat n_i-1) \rangle.
\end{align}

The data for the Bose-Hubbard model is shown as circles in Fig.~\ref{fig:vr_vs_Eint}. The interaction quench from $U/J=\infty$ to $U/J<\infty$ causes the dynamical formation of
doublons (there were none in the initial state). The trap opening induces a decrease of
$E_{\rm int}$ towards the asymptotic value. We do not reach the asymptotic regime in our simulations, but we choose a time large enough to capture most of the decay of $E_{\rm int}$. 
The results in Fig.~\ref{fig:vr_vs_Eint} suggest that for large $U/J$, the asymptotic radial velocity is indeed primarily a
 function of the interaction energy that is generated due to the interaction quantum quench over the first tunneling times~\cite{Enss2012,Bauer2015}.
There is a noticeable and expected additional $U$-dependence (see the inset where we plot $v_r$ versus  $E_{\rm int}/U$). This results from (i) the fact that doublons are only well-conserved objects for $U\gg W$ and (ii) that doublons expand with a nonzero velocity for finite $U\sim W$. The argument of Ref.~\cite{Sorg2014}, which explains the expansion velocities at large $U/J$, assumes immobile doublons (and higher site occupancies) on the relevant time scales, which is only correct for $U\gg W$.

The interaction energy is only a proxy for the actual heavy objects involved in the dynamics, in particular, since $E_{\rm int}$ still undergoes a slow decrease beyond the 
times reached in the simulations. 
A more rigorous argument is to relate $v_r$ to the overlap of the initial state with bound states (see~\cite{Boschi2014} for the two-body case) in the integrable 1D FHM model,
in extension of the approach taken in~\cite{Mei2016}.

\end{document}